# The Aging Metabolite Database with Summary Statistics

Delilah Milner, Oliver Fiehn

## Table of Contents

## 1.1 Abstract

Thousands of metabolomic papers are published each year, creating challenges for scientists to combine results and yield (biochemical) conclusions that span across studies. Literature databases such as the Human Metabolome Database (HMDB) provide summaries of metabolite detections and relevance, but it does not focus on specific processes, such as 'aging.' Another database, MetaboAge, focuses specifically on how metabolite concentrations change with age. However, both databases can only search metabolites individually. Both databases lack an easily displayed overview of all the information provided, and they do not combine datasets to validate results across studies. This project aims to design a database from data extracted from the literature that cover studies across species, include user-friendly query options, and combine data to get statistically sound results. The Aging Metabolite Database summarizes aging-related information for almost 2,700 non-unique metabolites and was created from 104 publications. Information can be searched by sample types such as organ, sex, species, and metabolite identifiers such as InChIKey and PubChem CID. Searches may be done individually or as a batch. Additionally, a sub-database was created to focus on metabolite plasma levels in human aging and, including metadata from 27 papers. This resulted in 6,500 data points from 1,167 metabolites. 12 of these papers were selected for the summary results section of the database. Across these 12 studies, 53 common metabolites were reported and can be queried in 163 summary statistics, for example by age and sex. Summary statistics are provided for all aging and sex combinations. This information assists users to compare the results of their own age-related studies to published literature. Combining multiple datasets results in a more solid understanding of validated metabolic changes during the aging process.

## 1.2 Introduction

Metabolomics studies enumerate changes in metabolites that are associated with biological phenomena. However, the more metabolomic studies get published, the harder it gets for an individual researcher to integrate findings in a statistically significant way in a specific field such as aging research[1]. For example, when attempting to retrieve metabolites that are age-associated from public databases, PubMed gives 2,500 results, and Google Scholar lists over 500,000 results. Manually comparing results from these reports is inefficient, error-prone due to problems in matching synonyms, and statistically unsound because results must be harmonized to give quantitatively valid conclusions. This problem has led to a need for chemical databases that collect data from the literature and display the results concisely. Querying published research is essential because it can increase the validity of current work and highlight its reproducibility. In metabolomics, the more often a metabolite has been observed to be significant, the more relevant that finding becomes. An example of a database of this nature is HMDB which was manually curated and first published in 2007[2]. HMDB is currently on its 5th version and stores chemical information on 418 metabolites that have been shown to increase or decrease with age[3]. The aging information provided includes the metabolite, the age that was tested, how the concentration of these metabolites changes with age, the organ tested, and the referenced publication. However, the problem with HMDB is that it only focuses on human studies and does not include information from other species, such as Caenorhabditis elegans (C. elegans), mice, or naked mole-rats[4,5].

The MetaboAge database is another example of a manually curated database that is similar to HMDB but focuses more on aging[6]. It includes information such as the metabolite's description, synonyms, and InChIKey for over 1,500 non-unique metabolites. The literature used in MetaboAge was pulled from PubMed. Unfortunately, both databases are limited in how data is searched and represented. They only allow searches by metabolite name. Therefore, searches cannot be done by sample type or in batches, and the databases do not provide overview data, or perform statistics on the available literature. The project presented here aims to create the Aging Metabolite Database, which includes more data and dramatically improves usability compared to others in the field by including more features and summary statistics.

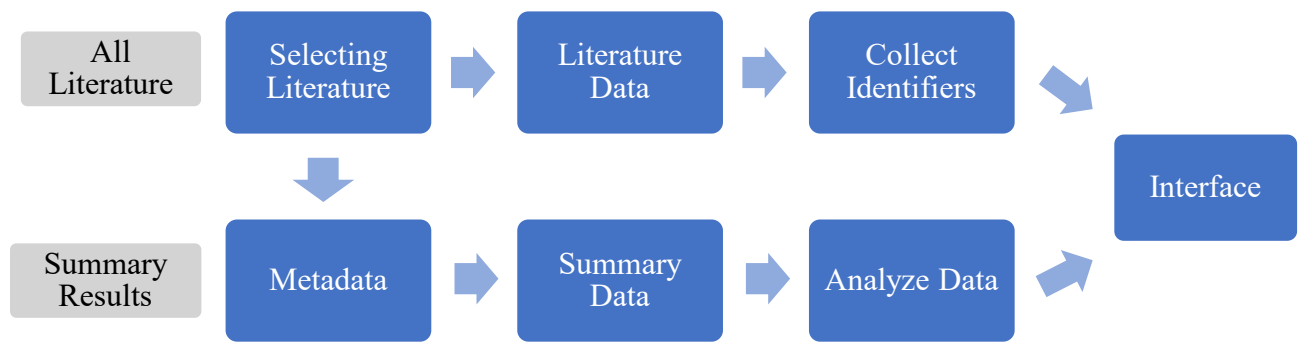

**FIGURE 1.**

The *Aging Metabolite Database* Workflow. Literature was selected based on meeting the keyword criteria. For all 104 papers, literature data, such as metabolite names and species, were collected. The resulting metabolite names were converted to other chemical identifiers using Chemical Translation Service (CTS) and PubChem, and additional information was added from the Metabolomics Workbench. For the summary results, 27 of the 104 papers were selected, and the means, standard deviations, and sample sizes were collected[7-33]. 12 of the 27 papers were used in the summary data[7,8,10,13,16,17,19,20,22,24,26,28]. PythonMeta was used to analyze the results. A user-friendly interface was created to present both datasets.

## 1.3 Methods

To create a more comprehensive aging database, searches go beyond PubMed and included articles from Google Scholar. The keywords included were "longevity", "metabolite", "aging", "healthy aging", and "centenarian". Only literature that contained the metabolite name, how the metabolite concentration changed with age, and the tested species and organ were added to the database. This filtering process resulted in 104 published papers. The metabolite information typically came from the titles, abstracts, results, and supplemental information. Almost 2,700 statistically significant metabolites were found.

After collecting the metabolite data, the Chemical Translation Service (CTS) and PubChem were used to convert the metabolite names to other chemical identifiers, such as InChIKey, KEGG, and LIPID MAPS[34-37]. The Metabolomics Workbench converted chemical names and InChIKeys into the standardized name, formula, exact mass, and subclass[38]. For metabolites missing results, they would be passed through the Metabolomics Workbench to collect more data.

To create the summary statistics, literature needed to include the means, standard deviations, and sample sizes. Data did not need to be considered significant to be included, as the summary result would determine significance[39]. This resulted in 27 papers studying humans from 2 organs and included over 6,500 data points from 1,167 metabolites[7-33]. 12 papers were selected for the summary statistics which includes 53 unique metabolites, and 167 comparisons, with 82 based on age[7,8,10,13,16,17,19,20,22,24,26,28]. The mean difference was used as the effect size for targeted metabolomics that used the same concentration units, while standardized mean difference was used for untargeted studies and studies that did not use the same units. Additionally, the fixed effect model was used for comparisons that contain only 2 studies, while the random effects model was used for comparisons that include 3 or more studies[40]. A user-friendly interface was created for the completed datasets using Dash in python[41]. The visualizations and data analysis used the Python libraries: Matplotlib, NumPy, PythonMeta, Plotly, and Pandas. A depiction of this process can be found in Figure 1.

**1.4 Results**

The all-literature dataset in the Aging Metabolite Database contains data from 100 published papers. The data includes 2,694 non-unique metabolites with 13 species and 25 organs tested. This is significantly more data than the 418 age-related metabolites found in HMDB and the 1,500 metabolites found in MetaboAge. HMDB also lacks data on other species which is important information to understand aging. Figure 2 shows that the most studied species were humans, naked-mole rats, and C. elegans, which means that there is valuable information missing from HMBD. Figure 2 also shows that the most studied organs were plasma and serum.

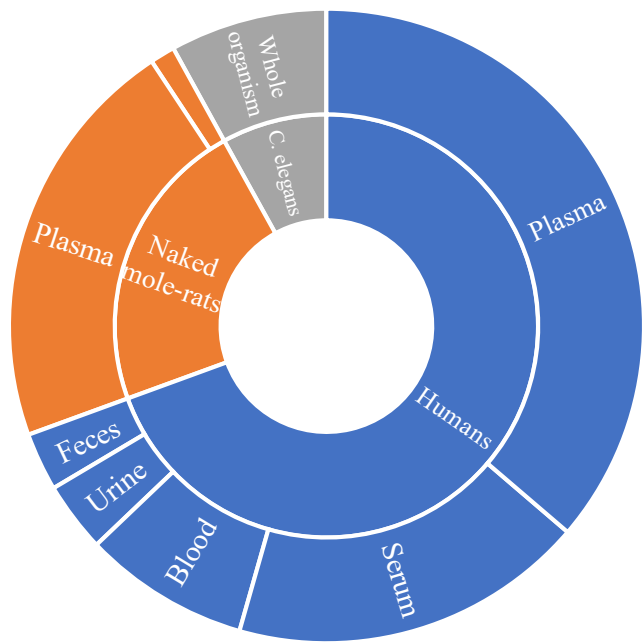

**Figure 2.**

Visualization of the number of metabolites found in different organs for the top-3 most reported species in aging research. The number of metabolites is shown by the relative size of the organs

      The all-literature dataset in the Aging Metabolite Database is organized by tables, with rows as metabolites and columns detailing organs, sex, species, and direction of metabolite change by age, and columns including chemical identifiers and additional information such as standardized name. The data can be searched in multiple ways, making the database more versatile compared to other literature databases. For example, the database can be searched using a single search, a batch search, and frequency and can easily be downloaded as a CSV file. All inquiries can be made using any column, such as sex and species, which is unique to the Aging Metabolite Database. Inquiries can also be further filtered by any combination of data. The single search feature will return all values for that given search.

For example, if the species "C. elegans" were searched, there would be 178 rows. Batch searches can be used to find how many times multiple values have been referenced. Table 1 is an example of a batch search on species where the values on the left are the values that can be input by the user. On the right are the results for that search. The displayed results are listed by count from highest to lowest.

| Input values | | Output | Count |
|---|---|---|---|
| elegans | 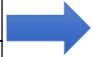 | Humans | 1605 |
| human | | Naked mole-rats | 500 |
| Naked mole-rats | | C. elegans | 178 |

**Table 1.**

User queries on species input (left) yield the number of metabolites reported to significantly change by age (right).

The frequency tab can be used to see which options are available. For example, selecting the species column yields a table showing the 13 species that were studied. Such query can be useful to see an overview of the collected data. Multiple columns can be searched in the frequency tab at once and the output results can be further filtered.

No other literature database enables the combining of multiple studies to create a mathematical meta-analysis comparing two different groups. This is beneficial as it can combine data obtained through both data-dependent and data-independent acquisition methods, as well as data collected across multiple instruments. For every combination, the meta-analysis results include the final effect size and the summary Z-score. For heterogeneity, it includes the $\tau^2$ value, $I^2$, and Q-statistic. The formulas for these calculations are found in Figure 3. The effect size is given by the difference between the two means.

a
$$\theta = \frac{\Sigma w_i \theta_i}{\Sigma w_i}$$

b
$$Z = \frac{\theta}{SE(\theta)}$$

c
$$\tau^2 = \frac{Q - (k-1)}{\Sigma w_i - (\Sigma w_i^2) / \Sigma w_i}$$

d
$$I^2 = \left(\frac{Q - (k-1)}{Q}\right) \cdot 100\%$$

e
$$Q = \Sigma w_i (\theta_i - \theta)^2$$

**Figure 3.**

The formulas used to calculate the summary results. (a). final effect size. $\theta$ is the mean difference and w is the weight of each study. (b), Z-score. SE represents the standard error (c), $\tau^2$. K represents the number of studies (d), $I^2$ (e), and Q-statistic.

    The meta-analysis dataset includes 6,500 data points and 1,167 metabolites, derived from 27 papers[7-33]. Due to the scarcity of data as not all literature analyzed the same metabolites, only 12 papers were retained for the meta-analysis calculations[7,8,10,13,16,17,19,20,22,24,26,28]. This resulted in 167 comparisons, including 82 comparisons based on age and 85 comparisons based on sex. Figure 4 and Table 2 shows an example of a specific comparison for tyrosine levels in human serum, testing 70-year-olds to 30-year-olds across three separate human cohort studies. When using the random effects model with mean difference of serum tyrosine levels as the effect size, this comparison resulted in a total sample size of 266 human participants. The overall result in Table 2 shows that 70-year-olds have 6.9 µM more serum tyrosine than 30-year-olds with a p-value < 0.1. The heterogeneity p-value 0.454 suggests there was no significant heterogeneity in the cohorts which is confirmed by an $I^2$ value of 0%

(for I² value ≤ 50%, studies are considered homogeneous). This example comparison shows that the new Aging Metabolite Database can perform statistically solid comparisons across multiple studies. The completed metabolite database can be found with the link below.

https://github.com/metabolomics-us/Delilah-AMDB

| Export | | | | | | |
|---|---|---|---|---|---|---|
| Tyrosine | Units: µM | Organ: Serum/Plasma | | | | |
| | | 70 year olds | | | 30 year olds | |
| StudyID | Mean | Standard Deviation | Sample Size ---- | Mean | Standard Deviation | Sample Size |
| Kouchiwa_et_al._2012 | 77.15 | 10.53 | 132.50 -- | 69.23 | 8.99 | 43.50 |
| Pitkänen_et_al._2003 | 70.00 | 16.00 | 24.00 -- | 64.50 | 12.59 | 24.00 |
| Mota-Martorell_et_al | 35.82 | 11.64 | 21.00 -- | 32.61 | 11.23 | 21.00 |
| | | | | | | |
| Effect measure: MD | Effect model: Random | Algorithm: IV | | | | |
| | | | | | | |
| Study ID | N | EffectSize [95%CI] | Weight(%) | | | |
| Kouchiwa_et_al._2012 | 176.00 | 7.917[4.699,11.136] | 72.85 | | | |
| Pitkänen_et_al._2003 | 48.00 | 5.5[-2.645,13.645] | 11.38 | | | |
| Mota-Martorell_et_al | 42.00 | 3.21[-3.707,10.127] | 15.77 | | | |
| Total | 266.00 | 6.9[4.153,9.647] | 100.00 | | | |
| 3 studies included | (N=266) | | | | | |
| Heterogeneity:Tau²=0.000 | Q(Chisquare)=1.59 | (p=0.454); | I²=0% | | | |
| Overall effect test: | z=4.92, p=0.000 | | | | | |

**Table 2.**

Summary statistic comparing serum tyrosine levels of 70-year-old to 30-year-old humans. Units are in µM.

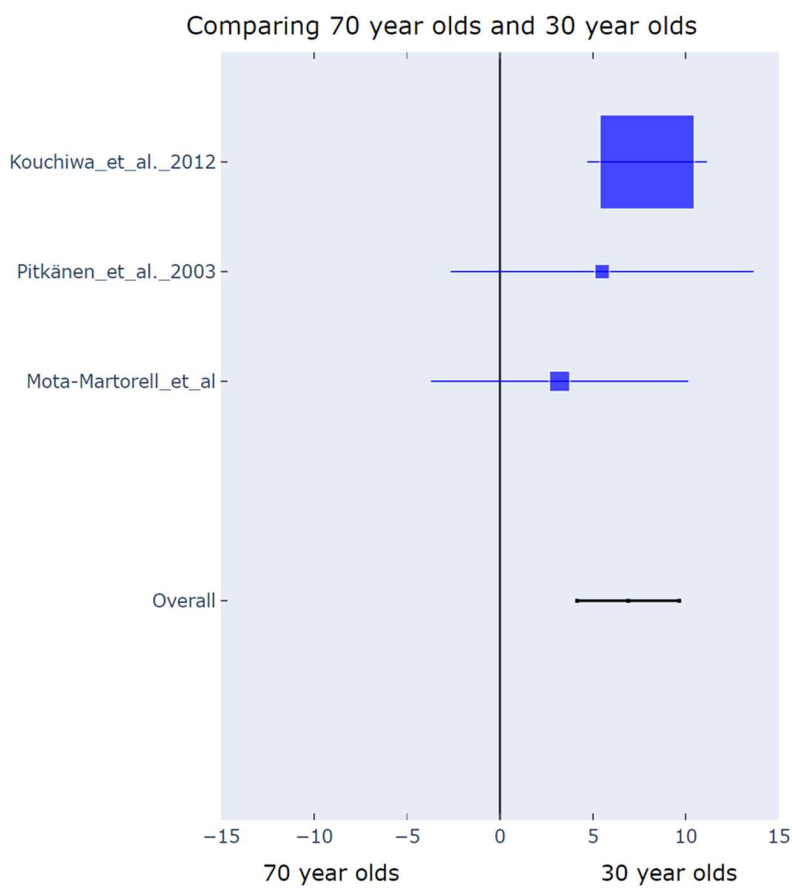

**Figure 4.**

Forest plot comparing serum tyrosine levels of 70-year-old to 30-year-old humans. Units are in μM.

## 1.5 Discussion and Conclusion

     The Aging Metabolite Database is a more versatile than previous databases for comparing metabolome studies, because it can combine data from both targeted and untargeted studies. Due to the scarcity of well-annotated metabolome data in reports on aging, most of the meta-analysis comparisons were confined to amino acids that are most frequently reported and most straightforwardly retrieved by synonym searching in study reports. The difficulties and inconsistencies in chemical synonyms used in metabolomic reports, and the refusal of authors and journals to use standardized identifiers such as International Chemical Identifier Keys, render our database not representative of the full human

metabolome. While there may be the possibility of automating the process of collecting and standardizing chemical names via large language models (LLM) and machine learning (ML) in the future, those technologies were not developed enough for this project. The first step would be training a model to identify appropriate literature. This project did not include studies investigating aging-related diseases or calorie restriction which a machine-based language model would find by using the PubMed Advanced Search Builder. Additionally, not all literature is public by machine-reading software, even for PubMed searches that limits full-text retrievals and instead, only allows full-abstract text mining. This deficiency means that many studies may still be unavailable to full-text mining, even if LLMs were better developed for chemical language, chemical structure graphs, and chemical synonyms. To improve the current stage of the Aging Metabolite Database would require retrieving more metabolomic data. This step would be equally challenging as some data that we included in the Aging Metabolite Database had to be manually extracted from embedded images and unlabeled data from the supplemental information associated with the published articles. The most straightforward way to automate literature databases would be for scientists to upload their data into repositories such as the MetabolomicsWorkbench or MetaboLights. Unfortunately, these databases also did not provide sufficient metadata (especially on details of biological study designs) to enable a comprehensive query of metabolome changes during aging. Until better repositories are created, literature databases will need to be created manually.